\documentstyle[prl,aps,preprint]{revtex}

\def\sqr#1#2{{\vcenter{\hrule height.#2pt\hbox{\vrule width.#2pt
height#1pt \kern#1pt \vrule width.#2pt}\hrule height.#2pt}}}

\newcommand{\be}{\begin{equation}}
\newcommand{\ee}{\end{equation}}
\newcommand{\ben}{\begin{eqnarray}}
\newcommand{\een}{\end{eqnarray}}
\newcommand{\bec}{\begin{center}}
\newcommand{\eec}{\end{center}}

\begin{document}
\preprint{gr-qc/9412046}

\draft
\widetext

\title{Classification of inflationary
Einstein--scalar--field--models via catastrophe theory}

\author{Fjodor V.~Kusmartsev}

\address{Institute for Solid State Physics, University of Tokyo,
        Roppongi, Minato-ku, Tokyo 106, Japan}

\author{Eckehard W.~Mielke, Yuri N.~Obukhov, and
Franz E.~Schunck\footnote{Electronic address: fs@thp.uni-koeln.de}}

\address{Institute for Theoretical Physics, University of Cologne,
D--50923 K\"oln, Germany}

\date{\today}

\maketitle

\begin{abstract}
Various scenarios of the initial inflation of the universe are
distinguished by the choice of a scalar field {\em potential}
$U(\phi)$ which simulates a {\it temporarily}
non--vanishing {\em cosmological term}.
Our new method, which involves a reparametrization in terms of the
Hubble expansion parameter $H$, provides a
classification of allowed inflationary potentials and of the stability
of the critical points. It is broad enough to
embody all known {\it exact} solutions involving one
scalar field as special cases. Inflation corresponds
to the evolution of critical points of some catastrophe manifold.
The coalescence of its nondegenerate critical points with the creation
of a degenerate critical point corresponds the reheating phase of
the universe. This is illustrated by several examples.
\end{abstract}
\bigskip

\narrowtext

{\bf Introduction}.
In the 80's, Guth \cite{guth} and Linde \cite{linde82} have
modelled an inflationary phase of the universe (cf.~\cite{sato}).
Scalar fields (Higgs, axion) are expected to generate, shortly after the
big bang, an exponential increase of the universe.
The so--called {\em graceful exit} to the Friedmann cosmos was partly
solved in the {\em new inflationary universe} \cite{albste}. In this model, the
scalar field is ruled by a slightly different self--interaction
potential which possesses a slow--roll part (a plateau)
of the potential (acting as a vacuum energy) which
dominates the universe at the beginning.
Power--law models were constructed which possess no
exponential but an $a(t) \sim t^n$ increase of the expansion factor of
the universe \cite{abbo,lucmat,barrow87}.
The {\em intermediate inflation} is merely
a combination of exponential and power--law increase
\cite{barrow90}. Further solutions were found in \cite{bar9394}.
Recently, by using the Hubble expansion parameter $H$ as a new ``time"
coordinate, we \cite{schmie} were able to derive the {\it general}
Robertson--Walker metric for a {\em spatially flat} cosmos.
Our formal solution for arbitrary $U(\phi)$ comprises all
previous exact solutions.

For a rather general class of inflationary models the Lagrangian density reads
\be
{\cal L} = \frac{1}{2 \kappa } \sqrt{\mid g \mid}
 \Biggl ( R
   + \kappa \Bigl [ g^{\mu \nu } (\partial_\mu \phi ) (\partial_\nu \phi )
   - 2 U(\phi ) \Bigr ] \Biggr )  \; , \label{lad}
\ee
where $\phi $ is the scalar field and $U(\phi )$ the self--interaction
potential. We use natural units with $c=\hbar =1$.
A constant potential $U_0= \Lambda / \kappa $ would simulate the
cosmological constant $\Lambda $. We do not separately discuss
non--minimally
coupled Jordan--Brans--Dicke type models \cite{brans} since they
can be reduced to (\ref{lad}) via the Wagoner--Bekenstein--Starobinsky
transformation \cite{wag,beken,kasp,galt}. Let us concentrate on the
{\em flat} Friedmann--Robertson--Walker cosmos
\be
ds^2 = dt^2 - a^2(t) \left [ dr^2 + r^2 \left (
       d\theta^2 + \sin^2 \theta d \varphi^2 \right ) \right ]
\; , \label{metric}
\ee
where $a(t)$ is the time--dependent expansion factor with the
dimension {\em length}.
Flat space is anyhow favored in the inflationary scenario.
The scalar field depends only on the time $t$, i.e.~$\phi = \phi(t)$.

{\bf The Friedmann evolution equations}.
Let us assume that $a(t) \neq 0$, such that we can express our equations
completely \cite {schmie} in terms of the Hubble expansion rate
$H := \dot a(t)/a(t)$.
Only the diagonal components of the Einstein equation are non--vanishing. The
$(0,0)$ component involving the density $\rho $, reads
\be
3 H^2 = \kappa \rho
= \kappa \left ( \frac{1}{2} \dot \phi^2 + U \right )  \; . \label{1}
\ee
It describes the {\it conservation of the energy}. The $(1,1)$,
$(2,2)$, and $(3,3)$ components are given by
\be
2 \dot H + 3 H^2 = - \kappa p
= -\kappa \left ( \frac{1}{2} \dot \phi^2 - U \right )  \; , \label{2}
\ee
where $p$ is the pressure generated by the scalar field.
The resulting Klein--Gordon equation is
\be
\ddot \phi = - 3 H \dot \phi - U'(\phi )
\; , \label{scalar}
\ee
which is, after multiplication by $\dot \phi $,
\be
\frac {1}{2} ((\dot \phi)^2) \dot{} = - 3 H (\dot \phi )^2
 - \dot U \; . \label{scatra}
\ee
By linear combination of (\ref{1}) and (\ref{2}) we find the
{\em autonomous nonlinear system}
\ben
\dot H & = & \kappa U(\phi ) - 3H^2 =: V(H,\phi ) \; , \label{doth} \\
\dot \phi & = & \pm \sqrt {\frac {2}{\kappa }}
  \sqrt{3H^2 - \kappa U(\phi )}
 = \pm \sqrt {-\frac {2}{\kappa } V(H,\phi )}
 \; .\label{dotphi}
\een
The function $V(H,\phi )$ will turn to be the ``height function" in
Morse theory \cite{miln}. Observe that (\ref{dotphi}) is, in view of
(\ref{1}) and (\ref{2}), {\em a first integral} of (\ref{scalar}).
Moreover, $V\le 0$ in order to avoid scalar ghosts. For
the metric (\ref{metric}), the Lagrangian density (\ref{lad}) reduces
\be
{\cal L} = - \frac {3}{\kappa } \dot a^2 a + \left [
 \frac {1}{2} \dot \phi^2 - U(\phi ) \right ] a^3 \; .
\ee
Since the shift function is normalized to one for the metric
(\ref{metric}), the canonical momenta are given by
$p=\partial {\cal L}/\partial \dot a=-6Ha^2/\kappa $ and
$\pi = \partial {\cal L}/\partial \dot \phi = a^3 \dot \phi $.
[Incidentally, this
suggest to take the volume $a^3$ as a generalized coordinate and use
$P=\partial {\cal L}/\partial (a^3 \dot) = -2 H/\kappa $
as new momentum.]
The Hamiltonian or ``energy function'' is given by
\be
E = \Biggl [\frac{1}{2 } \dot \phi^2 + \frac{1}{\kappa }
    V(H,\phi ) \Biggr ] \, a^3
\ee
and vanishes for all solutions.

{\bf Catastrophe of Whitney manifold and the critical points of the evolution}.
In the phase space \cite{piran}, the equilibrium states of the system
(\ref{doth}) and (\ref{dotphi}) are given by the
constraint $\{ \dot H,\dot \phi \}=0$. The critical or equilibrium points,
respectively, of this system are determined by
$V(H_c,\phi_c)=0$. This constraint is {\em globally} fulfilled by
$\kappa U(\phi ) = 3H_{\Lambda }^2$, where the Hubble expansion rate
is constant,
i.e.~$H_{\Lambda }=:\sqrt{\Lambda /3 \,}$. For $\dot \phi=0$ and
$\Lambda \neq 0$, we obtain the de Sitter inflation with
$a(t) = a_0 \exp (\sqrt{\Lambda /3 \, } t)$.

The Jacobi matrix $J$ of the system (\ref{doth}) and (\ref{dotphi}) is
given by
\be
J = \pmatrix {-6 H & \kappa U' \cr
       \pm 6 H (- 2 \kappa V)^{-1/2}
     & \mp \kappa U'(- 2 \kappa V)^{-1/2} \cr }
\; ,
\ee
where $U'=dU/d\phi $. Since $\det J = 0$, the system is degenerate.
For the analysis of stability, it suffices therefore to consider
only (\ref{doth}) and later to reconstruct $\phi$.  
Since $H$ and $\phi$ are independent variables, we can introduce
the non--Morse potential $W(H,\phi)$, which is defined via
$V:=-\partial W/\partial H$ and analyse the system with the aid of catastrophe
theory. From (\ref{doth}) we obtain
\be
W(H,\phi) = H^3 - \kappa U(\phi ) H + C(\phi) \; ,
\label{whitney}
\ee
where $C$ is an arbitrary function of $\phi $. The function $W$
is already in canonical form in $H$--space, and belongs to
a Whitney surface or to the Arnold singularity class $A_2$
(cf.~\cite{kusm}). That means that our catastrophe manifold here is the
Whitney surface. This manifold has only one control parameter, which is
here given explicitly as the potential $U$.
Thus, an evolution of critical points are determined via
the values of the potential $U$. Let us analyse the types of
critical points at different fixed values of the control parameter $U$.

If $U_c :=U(\phi_c)<0$, the equation has no stable critical
point due to the shape of the Whitney surface. However if
$U_c>0$, there are two critical points: stable
at $H_c=\sqrt{U_c/3}$ and unstable at
$H_c=-\sqrt{U_c/3}$. For $H_c>0$, Walliser \cite{wall} comes to the same
conclusion.

The value $U_c=0$ is the {\em bifurcation point}. Provided
this is also an extrema of $V$, we necessarily have
$\partial V/\partial H \mid_c=-6H_c=0$,
$\partial V/\partial \phi\mid_c =\kappa U'_c=0$, and $\dot \phi_c=0$.
Thus, also the critical points of the Klein--Gordon equation are involved.
Hence, the Hubble parameter
has to vanish and $\phi_c$ is a double zero of
the potential $U$. The Hessian of (\ref{doth}) takes the form
\be
Hess (V) = \pmatrix {-6 & 0 \cr
           0 & \kappa U'' \cr } \; .
\ee
The sub--determinant of the Hessian is
$\Delta_{0}=\partial^2 V/\partial H^2=-6<0$
and
$\Delta_1= \det Hess (V)= (\partial^2 V/\partial H^2) \;
(\partial^2 V/\partial \phi^2)
- (\partial^2 V/(\partial H \partial \phi ))^2 = -6 \kappa U''$.
For a maximum of the potential $U$, i.e.~$U'=0$ and $U''<0$,
the function $V$ possesses a maximum; for a
minimum of the potential $U$ we find a saddle point for $V$.

Alternatively, we can investigate the non--Morse potential
(\ref{whitney}) alone, e.g.~for the {\em chaotic inflationary model} with
$U(\phi)=\phi^2/\kappa$
\be
W(H,\phi) = H^3 - \phi^2 H + C(\phi) \; , \label{whitchaot}
\ee
which has minimum and maximum at $H_c=\pm \phi_c/\sqrt{3}$
and a saddle point for $\phi_c=0$. This corresponds to the end of
inflation and a {\it reheating} of the universe \cite{kofm}.

In the case of {\em power--law inflation}, we have
$U(\phi)=(\exp\phi)/\kappa$ and therefore obtain
\be
W(H,\phi) = H^3 -  e^\phi H + C(\phi) \; ,
\ee
which has the critical points $H_c=\pm e^{\phi_c/2}/\sqrt{3}$
and a saddle point
for $\phi_c=-\infty $. The latter can be reached because of
$\phi = \ln (1/t)$ for $t \rightarrow \infty $.
In both models, the saddle point appears at that time of the reheating of
the universe.

The last example here should be taken from the {\em new inflationary
model} with $U(\phi)=[\phi^2 (\phi^2 -A)+D]/\kappa $, where $A,D$ are
two constants. The function
\be
W(H,\phi) = H^3 - [\phi^2 (\phi^2 -A)+D] H + C(\phi)
\ee
possesses minima and maxima at $H_c=\pm \sqrt {U(\phi_c)/3}$
and two saddle points at $\phi_c=\pm \sqrt{A/2}$, the zeroes of the
potential, if $D=A^2/4$. That means that a shift of the potential is
necessary, such that the zeroes of the potential are also its minima.

{\bf   The reheating phase of the universe}.
Since $U(\phi_c)$ corresponds to the {\it latent heat} of the universe in this
phase, we are now in the position to state the following:

{\bf Theorem:} {\em The critical points of the non--Morse potential
$W(H,\phi )$ determine the evolution in the inflationary phase.
Along the minima and maxima $H_c=\pm \sqrt{U(\phi_c)/3}$,
the inflaton moves from the slow--roll
to the hot regime. The saddle points of $W$, i.e.~more precisely, the minima of
$V$, determine the onset of reheating}.

The critical points are related to some turning
points of the evolutions. Since our potential $U$ or the field
$\phi $, respectively, depend on time, we obtain the following scenario:
The system starts from some initial conditions.
In the first stage of the inflation, the stable critical points
define the regime of the evolution, that is the system evolves towards
such points. In the second stage, because our control parameters
(here the potential $U$ or the field $\phi $, respectively) also depend on
time, the evolution moves along the critical points.
Consider, for example, the chaotic inflationary model in
Eq.~(\ref{whitchaot}): The system being initially
located in a minimum, evolves to the saddle point; physically, this
means the end of the inflation and a reheating of the universe.
In this way, we have derived a {\em universality class} for inflationary
models. This picture is very general, as we are going to show in the next
section by means of catastrophe theory.

{\bf Bifurcation and other catastrophes}.
For the analysis of critical points of multidimensional functions
(see \cite{kusm}) the Hessian of (\ref{whitney}) is usually
employed. It has the form:
\be
Hess (W) = \pmatrix {6 H & -\kappa U' \cr
        -\kappa U '
     & -H \kappa U''+C'' \cr }
\; .
\ee
Thus the classification of all critical points may be given by the
Whitney theorem \cite{kusm}. However,
the explicit form of (\ref{whitney}) allows to solve this problem
without an analysis of the Hessian.
The given form of $W$ suggests that we have here {\it umbilic catastrophes}.
There are only a few of them:
elliptic $D^-_4$ or hyperbolic $D^+_4$, second elliptic or second hyperbolic
($D^-_6$ or $D^+_6$, respectively), and symbolic umbilic $E_6$.
With the knowledge of the form of these catastrophes, which are
given for example in Ref.~\cite{kusm} one can completely predict the
evolution of the system. For the example of the chaotic
inflation, Eq.~(\ref{whitchaot}) describes an {\em elliptic umbilic}
catastrophe which has, in this special form, only one critical point.

For $\kappa U(\phi ) \neq 3H^2$, we find
$\{ \dot H, \dot \phi \} \neq 0$, which implies that the solutions
$\phi = \phi (t)$ and $H=H(t)$ are {\em invertible}, i.e.
$t = t(H) =\int\frac {dH}{\kappa \widetilde U - 3 H^2}$.
In contrast to the construction of Lidsey \cite{lid} in which the scalar
field itself is employed as a new time variable, our approach is also
valid for the end of inflation. Then we can write the
potential in (\ref{doth}) and (\ref{dotphi}) in the {\em reparametrized} form
\cite{schmie}
\be
U(\phi ) = U(\phi (t)) = U(\phi (t(H))) = \widetilde U(H)  \; . \label{uh}
\ee

The {\em reduced problem} is given by the one--dimensional equation
\be
\dot{H} = \kappa\widetilde{U}(H) - 3H^2 = g(H) \; ,
\label{one}
\ee
where $g(H)$ is the ``graceful exit function'' of Ref.~\cite{schmie}.

All solutions of this equation may also be classified with the aid of the
{\em catastrophe theory} \cite{kusm}. Eq.~(\ref{one}) has critical
points which are determined via the function $\widetilde W(H)$,
defined, analogously to $W$, by $g:=-d\widetilde W/dH$.
The simplest case arises when $\widetilde W(H)$ is exactly the
{\em Morse potential}, i.e.~$\widetilde W(H) = \lambda H^2/2$, i.e.
its critical points are nondegenerate.
In this simple case the
solution depends on the sign of $\lambda$. If $\lambda>0$, the
only possible stable state, to which the system evolves, is $H=0$.

For $\lambda<0$, we obtain $H\sim C\exp{\lambda t}$ and the system
has no stable critical points. In that case
$a\sim C_1 \exp({c \exp({\lambda t})})$. Then, the universe would be
expanding too fast.

The regimes of evolutions, or critical points, depend on the shape
of the potential $\widetilde W(H)$.
With the aid of the theory of singularities (in
that particular case with the aid of
{\em elementary catastrophe theory} \cite{kusm})
we may classify the inflation regimes.
If the potential $\widetilde W(H)$ is a smooth function
it belongs to the one of the Arnold classes $A_n$, where $n\geq2$.

The canonical form of the $n$--th class is $\widetilde W(H) =\lambda H^{n+1}$.
The co--dimension of that potential is equal to $n-1$,
that is, the number of {\em control parameters} is equal to $(n-1)$.

The critical points, which are
structurally stable correspond to the minima and maxima of the
polynomial
\be
\widetilde W(H)_{deform} ={\lambda} (H^{n+1} + \lambda_{n-1} H^{n-1}
+ \lambda_{n-2} H^{n-2} + ...+\lambda_0) \; .
\label{deform}
\ee
The system described by (\ref{one}) make an evolution to the minima
of $\widetilde W(H)_{deform}$.
In each of the minima, we have $\dot{H}=0$ and $H=const$.
The maxima of the potential (\ref{deform}) correspond to
unstable points for which the instability transforms the system to
its minima. With the change of the control parameters, some minima coalesce
with maxima or vice versa. As a result, the number of minima changes.
Each minimum is associated with a distinct constant value of the Hubble
expansion rate.

That is, each of the critical points corresponds to
a different de Sitter type inflation. However, as in the considered case
of the Whitney catastrophe the coalescence of these nondegenerate
critical points into degenerate one mean the
reheating phase of the universe. That is, locally, on the considered
complicated manifold, we have a Whitney sub--manifold. The inflationary
evolution and the picture of the creation of the reheating phase
is the same as it is described above.

However, there are a few exceptions related to the instabilities.
One of these instabilities for the non--Morse potential
we have already discussed above.
Asymptotically, for a large time scale, the
instabilities are defined by the leading term
of the polynomial $\widetilde W_{deform}(H)$.

Asymptotically, if $\lambda<0$ and $n\ge 0$, we may write for the
$n$--th Arnold class
\be
g(H) = -(n+1)\lambda H^n \; ,
\ee
such that (\ref{one}) yields the solution
\be
H = [C+(n^2-1) \lambda t]^{1/(1-n)}
\ee
where $C$ is some positive constant.
These instabilities have no physical meaning.

Among the {\em explicit models} which have been analyzed
in \cite{schmie}, we consider as an instructive example
$g(H)= -2 H^2/A^2 + 2 A^2 \lambda^2$
(in the notation of \cite{bar9394}). This corresponds again
to the Whitney catastrophe or to the second Arnold class $A_2$.
Therefore, we may expect the described above picture.
Let us show it explicitely.
 This ansatz leads to
\ben
\phi (t) & = & A \ln [\tanh (\lambda t) ] \; , \\
H(t) & = & A^2 \lambda \coth (2 \lambda t) \; , \\
a(t) & = & a_0 [\sinh (2 \lambda t)]^{A^2/2} \; , \\
U(\phi ) & = & A^2 \lambda^2 \left [ (3A^2-2)
 \cosh^2 \left (\frac {\phi }{A} \right ) + 2 \right ]
\een
as solution. Thus we have recovered one of the recent solutions
of Barrow \cite{bar9394}, for further details see \cite{schmie}.

Now we have our two possible descriptions of the investigations of the
critical points. The potential $U(\phi )$ has to be shifted to
\be
U(\phi ) = A^2 \lambda^2 (3A^2-2)
 \cosh^2 \left (\frac {\phi }{A} \right ) \; ,
\ee
so that the only critical point is at the origin $\phi =0$, which is a
minimum of $U$ and hence a saddle point of $W$.
Alternatively, we can use $g(H)$, which also has to be shifted into
\be
g(H)= -2 H^2/A^2 \; .
\ee
The non--Morse potential $\widetilde W=2 H^3/3 A^2$ possesses, too, a
saddle point at $H=0$. In the two equivalent descriptions we were
able to find that the saddle point corresponds to the reheating phase.

Furthermore, it is now possible to relate the ``slow--roll'' condition,
for the velocity of the inflationary phase, to the critical points
resulting from catastrophe theory. For inflation (with $\ddot a>0$) the
 two ``slow--roll parameters'' are given \cite{schmie}, in first order
approximation, by $\epsilon= -g/H^2$ and $\eta= -dg/d(H^2)\;$, where
$g$ is the ``graceful exit function" given in (\ref{one}). In this
reduced dynamics, they are effectively
determined by the first and second derivatives of the reduced
non--Morse function $W(H)$, i.e., more precisely by
$\epsilon= -(1/H^2)(dW/dH)$ and $\eta =(1/2H) d^2 W/(dH)^2$.
They will also determine the density fluctuations \cite{lidlyt}.

\acknowledgments
We would like to thank Peter Baekler, Friedrich W.~Hehl, and Yuval Ne'eman for
useful comments. The work of F.V.K.~has been supported
by the Ministery of Education, Science and Culture of Japan
he appreciates also the hospitality of ISSP.
E.W.M.~thanks D.~Stauffer for his leave of absence from Cologne, paid
by the Canada Council.
Research support for Y.N.O.~was provided by the Alexander von Humboldt
Foundation (Bonn) and for F.E.S.~by the Deutsche
Forschungsgemeinschaft, project He $528/14-1$.




\end{document}